\documentclass[sigconf]{acmart}  

\AtBeginDocument{%
  }

\acmConference[VRST '22]{28th ACM Symposium on Virtual Reality Software and Technology}{November 29-December 1, 2022}{Tsukuba, Japan}

\begin{document}

\title{Size Does Matter: An Experimental Study of Anxiety in Virtual Reality}


\author{Junyi Shen}
\affiliation{
  \institution{University of Tsukuba}
  \streetaddress{}
  \city{Tsukuba}
  \state{}
  \country{Japan}
  \postcode{305-0005}
}
\email{shen.junyi.xs@alumni.tsukuba.ac.jp}

\author{Itaru Kitahara}
\affiliation{
  \institution{University of Tsukuba}
  \streetaddress{}
  \city{Tsukuba}
  \state{}
  \country{Japan}
  \postcode{305-0005}
}
\email{kitahara@ccs.tsukuba.ac.jp}

\author{Shinichi Koyama}
\affiliation{
  \institution{University of Tsukuba}
  \streetaddress{}
  \city{Tsukuba}
  \state{}
  \country{Japan}
  \postcode{305-0005}
}
\email{skoyama@geijutsu.tsukuba.ac.jp}

\author{Qiaoge Li}
\affiliation{
  \institution{University of Tsukuba}
  \streetaddress{}
  \city{Tsukuba}
  \state{}
  \country{Japan}
  \postcode{305-0005}
}
\email{li.qiaoge@image.iit.tsukuba.ac.jp}

\renewcommand{\shortauthors}{Shen et al.}

\begin{abstract}
The emotional response of users induced by VR scenarios has become a topic of interest, however, whether changing the size of objects in VR scenes induces different levels of anxiety remains a question to be studied. In this study, we conducted an experiment to initially reveal how the size of a large object in a VR environment affects changes in participants' (N = 38) anxiety level and heart rate. To holistically quantify the size of large objects in the VR visual field, we used the omnidirectional field of view occupancy (OFVO) criterion for the first time to represent the dimension of the object in the participant's entire field of view. The results showed that the participants' heartbeat and anxiety while viewing the large objects were positively and significantly correlated to OFVO. These study reveals that the increase of object size in VR environments is accompanied by a higher degree of user's anxiety.
\end{abstract}



\maketitle

\section{Introduction}
The emotional response is one of the most important factors in user experience study of virtual reality products. Accordingly, research is actively being conducted on how VR environments affect users' emotional responses, particularly anxiety and fear.

\begin{table*}[t]
\caption{Dimensions of each large object}
\label{tab:1}
\resizebox{1\textwidth}{!}{
\begin{tabular}{ccccccccccccccc}
\hline
Large object             & Small cuboid & Small sphere & Small cube & Woman  & Tower  & Man    & Large cuboid & Buildings & Ferris wheel & Tree    & Large cube & Camera  & Large sphere & Planet  \\ \hline
OFVO                     & 0.42\%       & 0.94\%       & 1.21\%     & 3.92\% & 4.04\% & 4.55\% & 8.97\%       & 10.59\%   & 11.53\%      & 14.06\% & 15.90\%    & 20.97\% & 24.24\%      & 24.24\% \\
Vertical viewing angle   & 21.27°       & 21.27°       & 21.97°     & 73.30° & 64.34° & 73.13° & 75.76°       & 74.70°    & 65.74°       & 90.18°  & 75.94°     & 84.73°  & 90.18°       & 90.18°  \\
Hotizontal viewing angle & 11.21°       & 21.77°       & 24.79°     & 56.41° & 77.32° & 65.90° & 80.37°       & 87.94°    & 94.95°       & 91.88°  & 111.38°    & 127.58° & 104.13°      & 104.13° \\ \hline
\end{tabular}}
\end{table*}

Picture size has been found to have a significant effect on emotional perception. Bigger pictures elicit greater emotional arousal than smaller pictures \cite{cite6}. However, it is still unknown whether bigger 3D objects in virtual reality environment could trigger higher level of anxiety than smaller objects. We propose a hypothesis that the increased size of objects in VR scenes triggers higher levels of user anxiety.

In this study, we use a VR experiment to initially reveal how the size of objects in a VR environment affects people's anxiety level and heart rate. We investigated the subjective units of distress scale (SUDS)\cite{cite12}through experiments in which users observe virtual objects with varied sizes.

\section{Methods}
38 participants (15 of which are female) volunteered to participate in this study. The participants are aging from 22 to 31 (\emph{M} = 25.84, \emph{SD} = 3.33). All participants had normal or corrected normal visual acuity. The virtual scenario was rendered on a standard windows PC and displayed on a Head-Mounted Display (HMD) HTC Vive Pro Eye. We applied the NISSEI HR-70 watch type photoelectric expression pulse monitor PulNeo to record the heart beats of participants.

\begin{figure}[t]
  \centering
   \includegraphics[width=\linewidth]{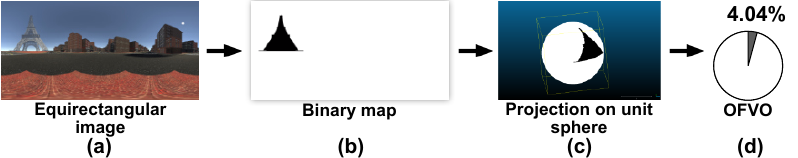}
   \caption{Calculation of omnidirectional field of view occupancy (OFVO)}
   \label{fig:4}
\end{figure}
In our experimental setup, participants could rotate their heads at will to observe the complete contour of the large object. Thus, we designed a new method for quantifying the object's size in the participant's field of view called omnidirectional field of view occupancy (OFVO) for this task, supplemented by the traditional horizontal and vertical viewing angles as the evaluation criteria for quantifying the size of the object. Since each participant had the same initial position in the VR scene, and we asked the participants to sit on the chair and not to move, the view observed by the participants could be approximated as a fixed omnidirectional image. OFVO is calculated as the proportion of the entire picture viewed by the participant that is occupied by the large object. We first captured the omnidirectional view observed at the subject's location in the VR scene in an equirectangular projection (Fig.~\ref{fig:4}a). Then, we manually binarized the omnidirectional image for each scene (Fig.~\ref{fig:4}b) and projected it onto a unit sphere model (Fig.~\ref{fig:4}c). Finally, we calculated the ratio of the projection of the large object on the sphere with respect to the entire sphere (Fig.~\ref{fig:4}d). We used Fibonacci lattice to sample the points on the sphere approximately equally spaced \cite{fibonacci2}. The number of points falling on the projection of the object on the sphere is then counted and divided by the total number of points to obtain the OFVO value. 
\begin{figure}[t]
  \centering
   \includegraphics[width=\linewidth]{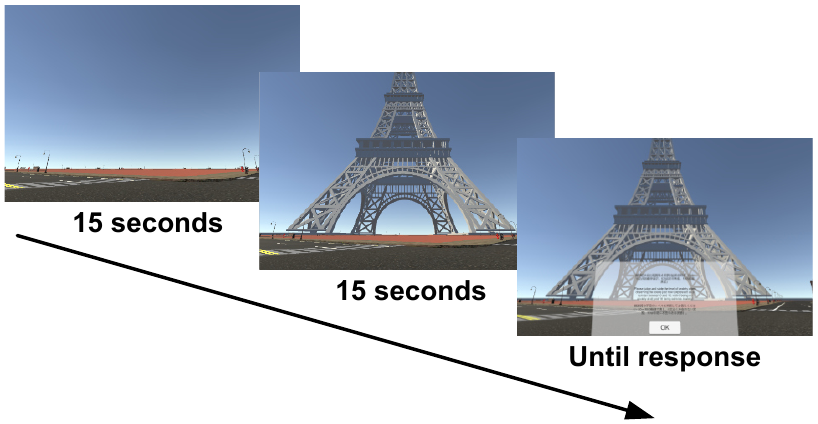}
   \caption{Procedure for each experiment trial}
   \label{fig:3}
\end{figure}

We used a city scene in unity. Our large objects are divided into 14 types. For each object, the OFVO and vertical and horizontal degrees of each object are listed in Tab.~\ref{tab:1}. The color of the cubes, cuboids and spheres are the same (grey, 158.17 $cd/m^2$, $x$ = 0.334, $y$ = 0.349). Large and small cubes, cuboids, and spheres are only different in size, and the shape of the 3D geometries are the same.

During the experiment, the participants used the controller to perform the experimental steps according to the text on the HMD screen. Each participant completed 14 trials, and in each trial, a large object that had not appeared in the previous trial appeared randomly in the field in front of the participant. Subjects were asked to sit in their seats and freely turn their heads to observe the current large object. After 15 seconds of free viewing, the participant reported the current SUDS on anxiety. While confirming that the response was complete and taking a 15-second break, the participant proceeded to the subsequent trial (Fig.~\ref{fig:3}). 

\section{Results}

\begin{figure}[t]
  \centering
   \includegraphics[width=\linewidth]{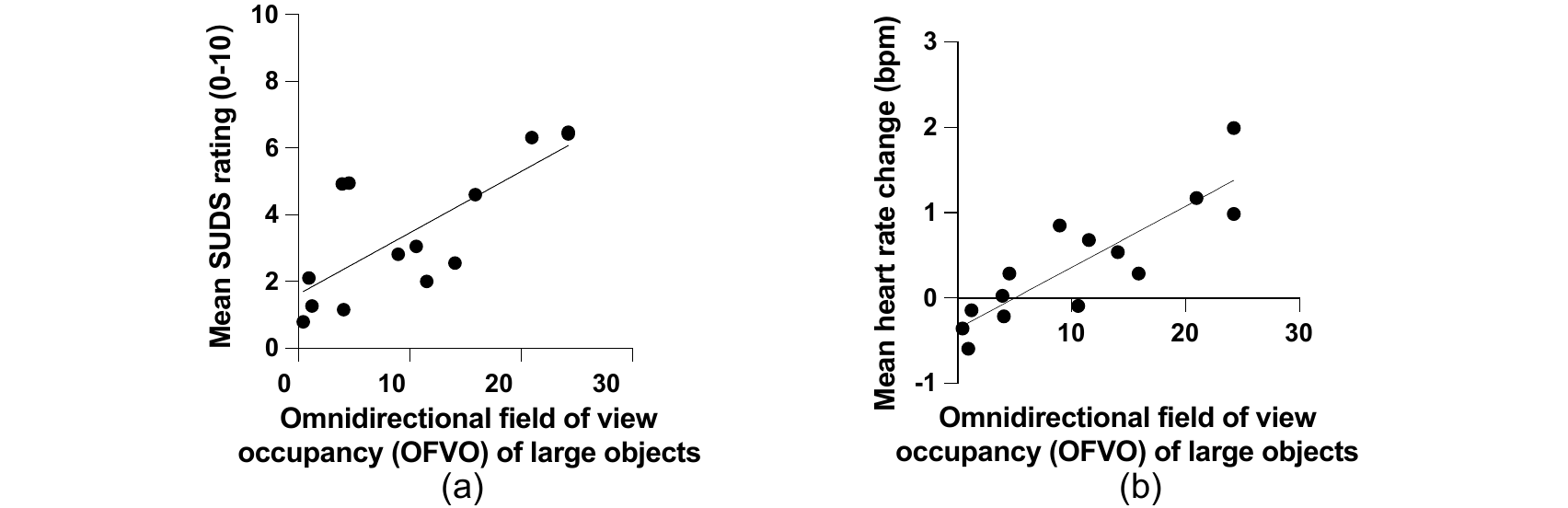}
   \caption{Response to the VR exposure in SUDS (a) and heart rate change (b)}
   \label{fig:2}
\end{figure}

38 participants completed all the measures of 14 trials. A series of correlation analyses were done to investigate the relationship between OFVO and anxiety. A significant correlation, $r$ = 0.76, $p$ = 0.002, between the SUDS and OFVO was found (Fig.~\ref{fig:2}a). 

After conducting an RM one-way ANOVA (analysis of variance), we found that among the three small geometric objects (cube, sphere and cuboid), although the sphere did not have the highest OFVO, participants reported significantly higher anxiety rating about the sphere ($M$ = 2.11, $SD$ = 2.07) than cube ($M$ = 1.26, $SD$ = 1.48) and cuboid ($M$ = 0.79, $SD$ = 1.13, $F$ (2, 111) = 6.35, $p$ = 0.002, $\eta^2$ = 0.103). There was no significant difference between the anxiety level of the small cube and the small cuboid ($p$ = 0.097). There were no significant differences between the large planet and the large gray sphere in SUDS ($p$ = 0.82). The texture image of the planet did not cause higher anxious. 



Participants' mean heart rate data at the last second of the 15-second rest period was recorded as a baseline. Participants' mean heart rate change relative to baseline in each scene were statistically analyzed. We found a significant correlation between the participant's heart rate changes and OFVO ($r$ = 0.86, $p$ < 0.001)(Fig.~\ref{fig:2}b).

\section{Conclusion}
In this study, we investigated how large objects in VR scenes affect people's anxiety and heart rate change. We proposed a new measurement to describe the size of objects in VR environments: OFVO, and discovered a positive correlation between OFVO and anxiety. We infer that in VR scenarios, objects of larger size trigger greater anxiety in users compared to objects of smaller size. The objects used in this experiment were not consistent in appearance except for size, which is a limitation of the study design. Manipulating the size of the same object is one way to improve the experimental design. This study has important implications for future research and VR product design on the perception and emotion of large-sized objects in VR environments.

\bibliographystyle{ACM-Reference-Format}
\bibliography{reference}

\end{document}